\documentclass[aps,showpacs,floatfix]{revtex4}
\usepackage{graphics}
\usepackage{dcolumn}
\newcommand{\bea}{\begin{eqnarray}}
\newcommand{\eea}{\end{eqnarray}}
\newcommand{\be}{\begin{equation}}
\newcommand{\ee}{\end{equation}}

\def\be{\begin{eqnarray}}
\def\ee{\end{eqnarray}}
\def\bd{\begin{displaymath}}
\def\ed{\end{displaymath}}

\def\PR{Phys. Rev. }
\def\PL{Phys. Lett. }
\def\NP{Nucl. Phys. }

\begin{document}
\title{Neutron Dripline in Odd and Even Mass Calcium and Nickel Nuclei}
\author{Madhubrata Bhattacharya}
\author{G. Gangopadhyay}
\email{gautam@cucc.ernet.in}
\affiliation{Department of Physics, University College of Science,
University of Calcutta\\
92, Acharya Prafulla Chandra Road, Kolkata-700 009, India}

\begin{abstract}
Neutron rich Ca and  Ni nuclei have been studied in spherical Relativistic 
Mean Field formalism in co-ordinate space. A delta interaction has been 
has been adopted
to treat the pairing correlations for the neutrons. Odd nuclei have been treated in the blocking 
approximation. The effect of the positive energy continuum and the role
of pairing in the stability of nuclei have been investigated using the
resonant-BCS (rBCS) approach. In Ca isotopes, 
$N=50$ is no longer a magic 
number while in Ni nuclei, a new magic number emerges at $N=70$. There is 
a remarkable difference in the relative positions of the drip lines for 
odd and even isotopes. In Ca isotopes, the last bound even and odd nuclei 
are found to be $^{72}$Ca and $^{59}$Ca, respectively. In Ni isotopes, the
corresponding nuclei are $^{98}$Ni and $^{97}$Ni, respectively. 
The origin of this difference in relative positions of the dripline in even 
and odd isotopes in the two chain is 
traced to the difference in the single particle level structures and consequent
modification in the magic numbers in the two elements. Pairing interaction 
is seen to play a major role. The effect of the width of the resonance states on pairing has also been investigated.
\end{abstract}
\maketitle


In recent years, it has been possible to populate and study a number of neutron 
rich nuclei from fusion evaporation reactions using radioactive ion beams 
as well as from fission fragment studies. The last bound neutrons
in such a nucleus may lie very close to the continuum and 
the effect of the positive energy continuum on the structure of such nuclei 
should be studied carefully. Another very important aspect of nuclei 
near the drip line is the additional stability provided by the pairing 
interaction. It is important to study the 
effect of pairing by studying the odd mass nuclei along with the
even mass ones. Although a number of such studies has 
been undertaken in lighter mass regions, nuclei in the medium and the 
heavy mass regions have not yet been studied in sufficient detail. In the 
present calculation, neutron rich Ca and  Ni nuclei have been studied using the 
Relativistic Mean Field (RMF) formalism in co-ordinate space.

RMF theory is a major tool of nuclear structure physics. Very often, the RMF 
equations are solved by expanding in a harmonic oscillator basis. However,
it is well known that the basis expansion approach using the harmonic 
oscillator basis cannot explain the density in halo nuclei near the drip line 
because of slow convergence in the asymptotic region. Nonrelativistic Hartree 
Fock Bogoliubov (HFB) and Relativistic Hartree Bogoliubov (RHB) methods in 
co-ordinate space 
have emerged as two very accurate approaches of treating the nuclei very close 
to the drip line. 

The effect of the states in the continuum has been incorporated in most of 
the calculations by solving the equations in coordinate space using the box 
normalization condition, thus replacing the continuum with a set of discrete 
positive energy states. However, in this case, the single particle energy 
levels depend on the size of the box chosen, leading to a rather unsatisfactory 
scenario. Nonrelativistic mean field equations involving continuum states have 
been solved with exact boundary conditions\cite{hfb1,hfb2} for zero range and
finite range pairing forces. RMF equations involving continuum states have also 
been solved\cite{rmfc,rmfc1} with exact boundary conditions. All these 
calculations have taken into account the effect of the width of the continuum 
levels also. For example, Cao and Ma \cite{rmfc1} have also compared their 
results with calculations assuming the continuum levels to be of zero width. 
We have also used the RMF calculation in co-ordinate space including the 
continuum states to study neutron rich even-even C and Be nuclei in 
detail\cite{C}. It has been observed that the results of this method and the 
more complicated RHB approach are in good agreement\cite{rmfc,C}.

In the present work we study the structure of neutron rich Ca and Ni nuclei 
beyond $^{48}$Ca and $^{68}$Ni, respectively, using the RMF formalism in 
co-ordinate 
space with exact boundary conditions, paying  particular attention to the odd 
mass nuclei. There have been numerous calculations on even-even neutron rich 
Ca and Ni nuclei up to the neutron drip line using relativistic and
nonrelativistic mean field approaches. We list only a few of them which have
been published recently. Nonrelativistic  Hartree Fock and Hartree Fock
Bogoliubov  approach have been utilized by Fayans {\em et al.}\cite{Cahf}
and Im {\em et al.}\cite{Cahf1} to study Ca isotopes. 
Relativistic density dependent Hartree Fock approach \cite{ddHF} as well as 
Relativistic Continuum Hartree Bogoliubov method (RCHB)\cite{RHB} have 
also been utilized to study neutron rich isotopes of
Ca and Ni\cite{RCHB1,RCHB}. Yadav {\em et al.}\cite{Yadav} have performed RMF+BCS 
calculation for Ca  and Ni isotopes by discretizing the continuum with box 
normalization.
The problem of wrong asymptotic behaviour in the harmonic oscillator basis
has been overcome using expansion in a Woods Saxon basis in a RMF calculation
\cite{WS}. The RCHB approach particularly gives a very good description of
the nuclei. Terasaki {\em et al.}\cite{Nihfb} have studied even-even 
Ni isotopes from the proton to the neutron drip line using HFB approach in 
three dimensions. However, we have come across no mean field calculation for 
odd Ni isotopes near the drip line. Even in Ca isotopes, the odd mass nuclei 
have not been studied in detail. The aim of the present work is to investigate 
the odd mass neutron rich Ni and Ca nuclei up to the drip line. 

The relativistic mean field theory is well known and readers are referred to
Refs. \cite{rmfrev} for details. We have used the force NLSH\cite{NLSH}.
We have assumed spherical symmetry for all the nuclei.
The BCS calculation has been performed adopting a delta interaction to
treat the pairing correlations 
between neutrons, {\em i.e.}
$V=V_0\delta(\vec{r}_1-\vec{r}_2)$. The usual BCS equations now contain 
contributions from the bound states as well as the resonant continuum. The 
equations involving these states have already been obtained\cite{hfb1,rmfc} 
and have been referred to as resonant-BCS (rBCS) equations.  
We have also included the effect of the width of the positive energy levels.
These equations have been solved in the co-ordinate basis on a
 grid of size 0.08 fm. The positive energy resonance solutions are
obtained using the scattering approach. All the negative energy states beyond 
N=20 and the positive energy states up to N=82  for which resonance 
solutions are available have been included in the rBCS calculation for studying 
the Ca isotopes. For the Ni isotopes, the corresponding numbers are N=28 and 
N=126, respectively.
We have assumed that beyond 20 fm, the effect of nuclear interaction vanishes.
We choose $V_0$= -350 MeV
for the strength of the delta-interaction. This value of the strength has
been used by Yadav {\em et al.}\cite{Yadav} in Ca nuclei. 
In odd nucleus the last odd nucleon
breaks the time reversal symmetry of the system. However, it is well known that 
the bulk quantities like binding energy or radii are not affected by the 
breaking of the symmetry. Since we are interested only in these quantities, 
we have used the blocking approximation to study odd mass nuclei.

We have calculated the binding energy corresponding to the different
levels in  odd nuclei. Although the ground state spin
parity is unknown in most of the nuclei, we find that, except for 
$^{75}$Ni, the spin parity has been correctly predicted wherever
the ground state is unambiguously known.
In  Fig. 1 we plot the binding energy values of the all the nuclei studied
in the present work which are stable against neutron emission.
The experimental or empirical values are from Ref. \cite{mass}. One can 
see that the theoretical values agree with experiment. In Figs. 2 and 3, 
we have plotted the one neutron separation energies for Ca and Ni isotopes,
respectively. The one neutron separation energy has been defined as
$S_n(Z,N)={\rm B.E.}(Z,N)-{\rm B.E.}(Z,N-1)$.
In Ni nuclei, the theoretical results are very close to experimental 
measurements. For the Ca isotopes, the agreement is slightly poorer. 
The calculated results for two neutron separation energy for the even mass 
nuclei are very close to the RCHB results obtained
by Zhang {\em et al.}\cite{RCHB}. 

One of the most interesting differences between the two chains studied is
the difference in the positions of the dripline for even and odd isotopes.
For Ca isotopes, the last odd mass nucleus stable against neutron emission
is $^{59}$Ca while the corresponding even mass isotope is $^{72}$Ca. On
the other hand, the corresponding nuclei in Ni  are $^{97}$Ni and $^{98}$Ni, 
respectively.  In other words, all the Ni isotopes are stable against
neutron emission up to $N=70$ beyond which neither the even nor the odd 
isotopes are stable for this element. For Ca the odd isotopes are
stable below $N=40$ while stable even isotopes range up to $N=52$.

To understand this dramatic difference we next study the single particle 
neutron levels. In Fig. 4  we have plotted the single particle 
energy levels near the Fermi level for even-even Ca  nuclei. 
 The Fermi level is indicated by the solid line. Near
the drip line the Fermi level rises slowly and finally becomes positive beyond
$^{72}$Ca. 
The states $1g_{9/2}$ and $2d_{5/2}$ start as positive energy states but
as the neutron  number increases they become loosely bound states.
Because of the absence of the centrifugal barrier,  the level
$3s_{1/2}$ does not have a resonant solution. The levels $3s_{1/2}$, 
$2d_{5/2}$ and $2d_{3/2}$ lie very close to each other and are close to the 
$1g_{9/2}$ level in $^{66-72}$Ca. The shell closure is taken to be the 
neutron number for which the pairing energy vanishes.
We find that in Ca isotopes, no shell closure 
is observed beyond N=40 because of the modification in the energy levels. 
The Fermi level beyond $N=40$ increases very slowly and lies very close to the 
continuum. The single particle density near is large near it. 
Hence pairing correlations which bind the loosely bound nucleons can develop in 
even-even nuclei. As pointed out by Meng {\em et al.}\cite{Meng}
these conditions indicate that a neutron halo should develop in these nuclei.
In Fig. 5, we have plotted the neutron radii of the nuclei studied.
The sudden increase in radius is clearly observed.
Importantly, in Ca isotopes, the level $1g_{9/2}$ lies either in the
continuum or just bound around N=40. 
Hence the stability of the isotopes depends crucially on the 
pairing interaction and the contribution of the states in the continuum. 
Thus the odd mass Ca isotopes beyond $N=40$ are unstable against neutron 
emission. There is a similar situation in light nuclei. The dripline nuclei
$^{11}$Li and $^{14}$Be are known to have two neutron halos. It is known that
neither $^{10}$Li nor $^{13}$Be are bound. The pairing interaction stabilizes 
the last neutron pair so that the above two dripline nuclei are stable
against neutron emission. 

On the other hand, the Ni isotopes present a completely different 
picture.  The single particle levels are shown in Fig. 6. Here, the Fermi 
level is deeper and abruptly becomes positive beyond A=98. The levels 
3$s_{1/2}$, 1$g_{7/2}$,  2$d_{3/2}$, and 2d$_{5/2}$ lie very close to each 
other. All of them except the first start as positive energy 
states and become bound states at higher neutron number.  The intruder orbit 
$1h_{11/2}$ lies much higher in energy. This is a consequence of the quenching 
of the spin orbit splitting in neutron rich nuclei which has been obtained in 
many mean field calculations and also experimentally observed in light nuclei. 
Thus $N=70$ becomes a new magic number. Because the occupied levels are much 
deeper, one does not expect a halo like structure and the binding  of the odd
nucleus in this case does not depend overmuch on the pairing. Hence
we obtain both even and odd mass nuclei which are stable against particle
emission up to the new magic number N=70. The calculated neutron radius 
values are shown in Fig. 5. One can see that the neutron radius shows
a shell closure at $N=50$, and more importantly, at $N=70$.

It is interesting to study the effect of the width of the resonant states 
in the nuclei very close to the drip line. In Table I, we present the 
results for some of these nuclei. The width of different levels, the binding
energies and the neutron radii ($r_n$) for $^{70,72}$Ca and $^{96,98}$Ni 
are presented. Pairing correlations usually become stronger if the
width of the resonances be neglected\cite{hfb1,rmfc} thus increasing the 
total binding energy and decreasing the neutron radius.
One can see that as $^{98}$Ni is a closed shell nuclei, 
there is no effect of the resonance states beyond $N=70$ in the
binding energy and the radius.

To summarize, neutron rich Ca and Ni nuclei have been studied using
RMF formalism in the co-ordinate space. Exact boundary conditions have been
used to obtain the wave function of the states in the continuum. A delta 
interaction has 
been used for interaction between neutrons. Odd nuclei have been treated in 
the blocking approximation. The study of odd mass neutron rich nuclei in this
mass region reveals interesting features. There is a 
remarkable difference in the relative 
positions of the drip lines for odd and even isotopes in the two elements. 
In Ca isotopes, the last bound even and odd nuclei
are found to be $^{72}$Ca and $^{59}$Ca, respectively. In Ni isotopes, the
corresponding nuclei are $^{98}$Ni and $^{97}$Ni, respectively. 
The origin of this difference in the relative positions of the drip line in 
even and odd isotopes in the two chains is 
traced to the difference in the single particle level structures and the
modification in the magic numbers in the two elements. 
In Ca isotopes, $N=50$ is no longer a magic
number while in Ni nuclei, a new magic number emerges at $N=70$. The level 
density near the continuum is higher in Ca isotopes and pairing interaction
plays a crucial role in the their stability. In odd mass nuclei, the 
pairing interaction is weaker and not able to bind the neutrons in the halo 
orbitals. If one neglects the width of the resonance states, the pairing 
correlation increases.

\acknowledgments{This work was carried out with financial assistance of the
Board of Research in Nuclear Sciences, Department of Atomic Energy (Sanction
No. 2005/37/7/BRNS), Government of India.}

\newpage
\begin{table}
\caption{Width of different resonant levels, and binding energies and neutron 
radii for calculations involving width (RMFW) and ignoring width (RMFN).}
\begin{tabular}{c|c|c|cc|cc}\hline
Nucleus&level&Width&\multicolumn{2}{c|}{B.E.(MeV)}
&\multicolumn{2}{c}{$r_n$(fm)}\\
       &&MeV&RMFW&RMFN&RMFW&RMFN\\\hline
$^{70}$Ca&2d$_{5/2}$ &0.102&468.30&468.35&4.728&4.581\\
         &1g$_{7/2}$ &0.874&&&&\\
         &1h$_{11/2}$&1.524&&&&\\
         &2d$_{3/2}$ &0.593&&&&\\
\hline
$^{72}$Ca&2d$_{5/2}$ &0.013&468.35&468.47&4.994&4.710\\
         &1g$_{7/2}$ &0.448&&&&\\
         &1h$_{11/2}$&1.501&&&&\\
         &2d$_{3/2}$ &0.478&&&&\\
\hline
$^{96}$Ni&1h$_{11/2}$ &0.033&665.02&665.32&4.952&4.926\\
         &1h$_{9/2}$  &1.678&&&&\\
         &1i$_{13/2}$ &2.474&&&&\\
         &2$_{7/2}$  &2.289&&&&\\
\hline
$^{98}$Ni&1h$_{11/2}$ &0.018&667.07&667.07&4.965&4.965\\
         &1h$_{9/2}$  &1.403&&&&\\
         &1i$_{13/2}$ &2.446&&&&\\
         &2f$_{7/2}$  &2.000&&&&\\
\hline
\end{tabular}
\end{table}
\begin{figure}
\resizebox{8cm}{!}{\includegraphics{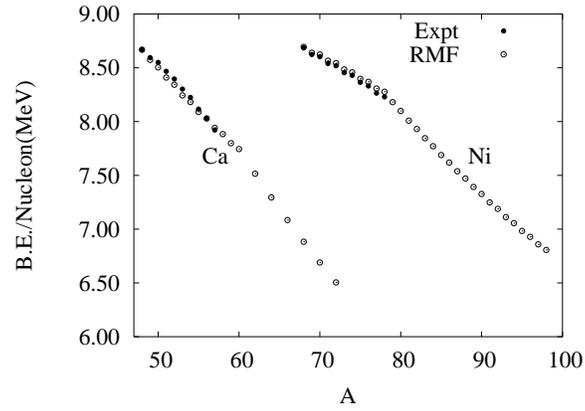}}
\caption{Experimental and calculated binding energy per nucleon in Ca and Ni nuclei.
\label{be}}
\end{figure}
\begin{figure}
\resizebox{8cm}{!}{\includegraphics{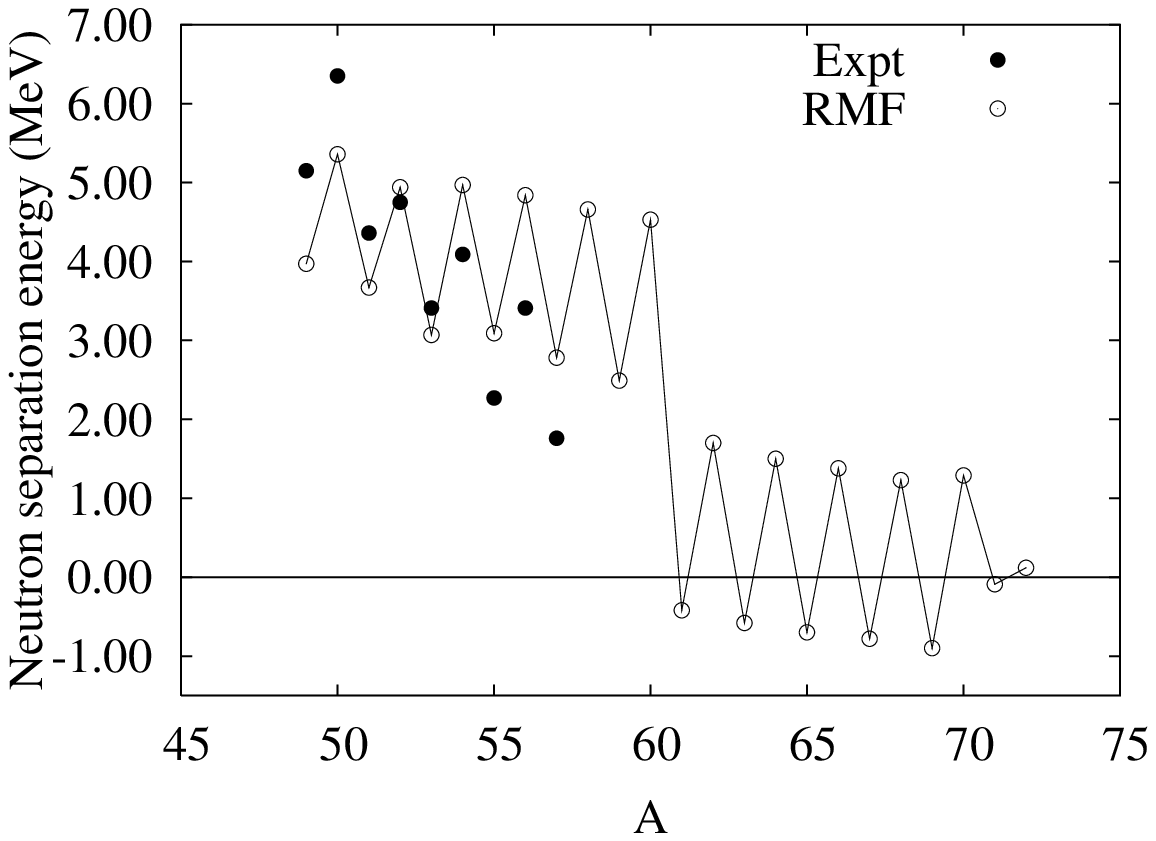}}
\caption{Neutron separation energy in Ca isotopes.\label{CaS2n}}
\end{figure}
\begin{figure}
\resizebox{8cm}{!}{\includegraphics{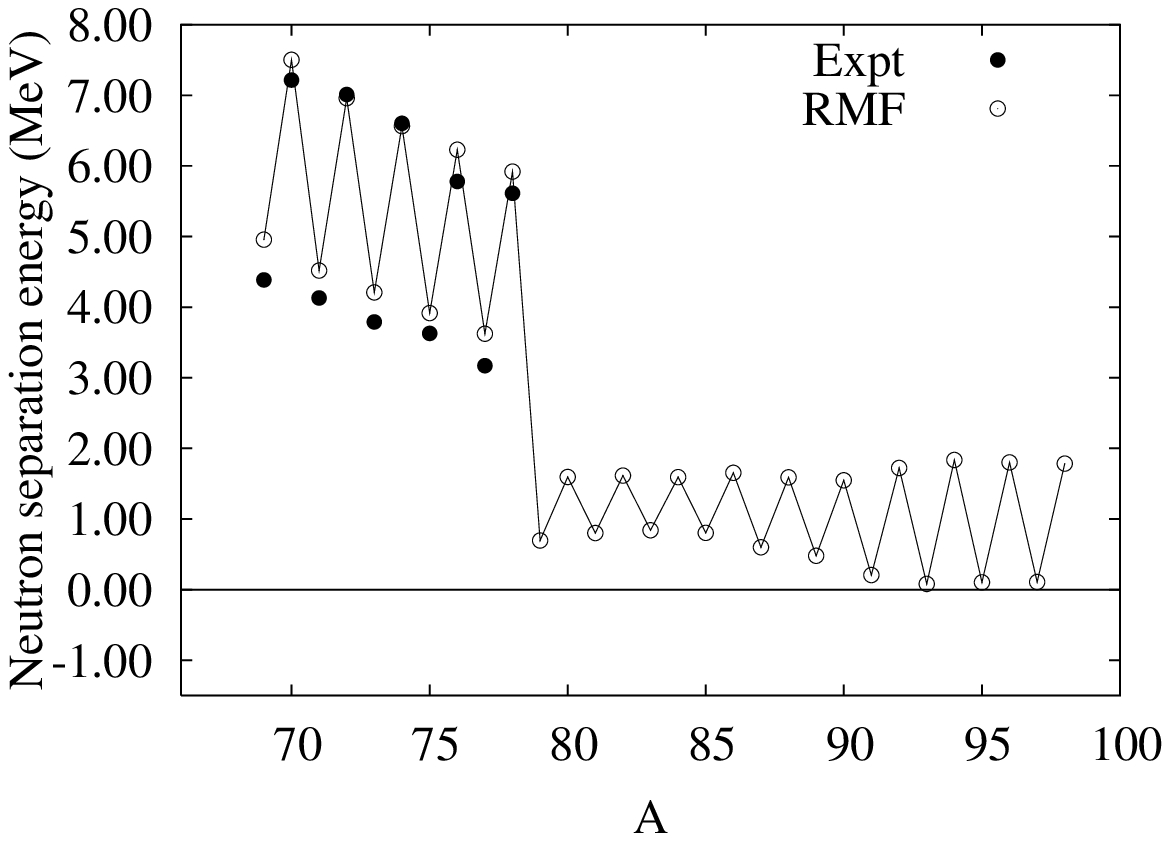}}
\caption{Neutron separation energy in Ni isotopes.\label{S2n}}
\end{figure}
\begin{figure}
\resizebox{8cm}{!}{\includegraphics{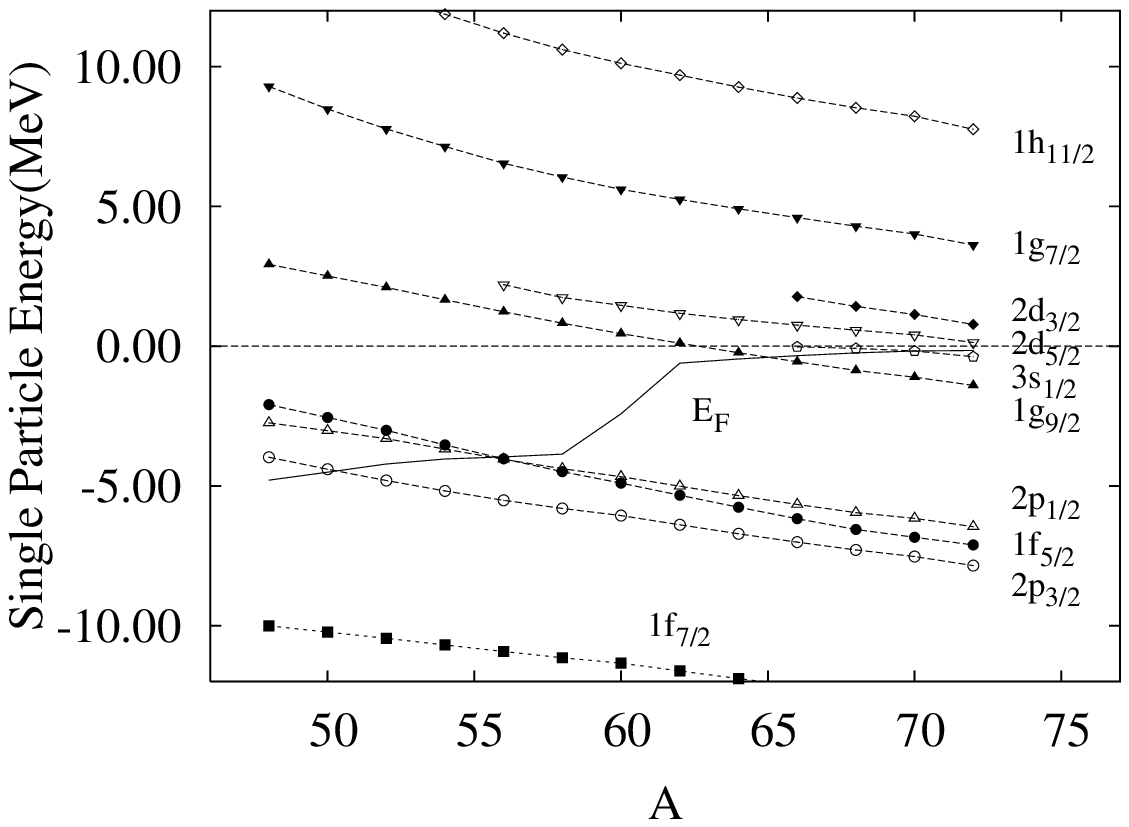}}
\caption{Single particle neutron energy levels in even
Ca isotopes. The Fermi level is shown by the continuous line.
\label{Calev}}
\end{figure}
\begin{figure}
\resizebox{8cm}{!}{\includegraphics{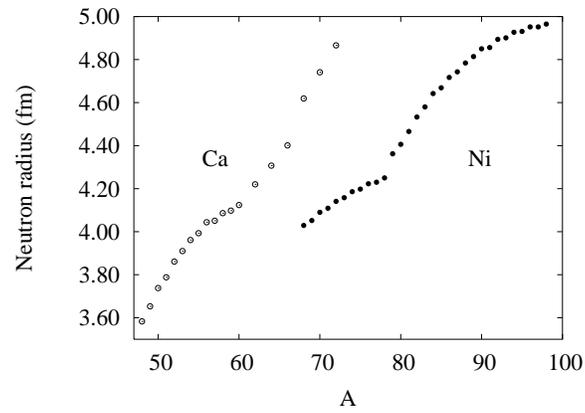}}
\caption{Neutron radius in neutron rich Ca and Ni nuclei.
\label{Carad}}
\end{figure}
\begin{figure}
\resizebox{8cm}{!}{\includegraphics{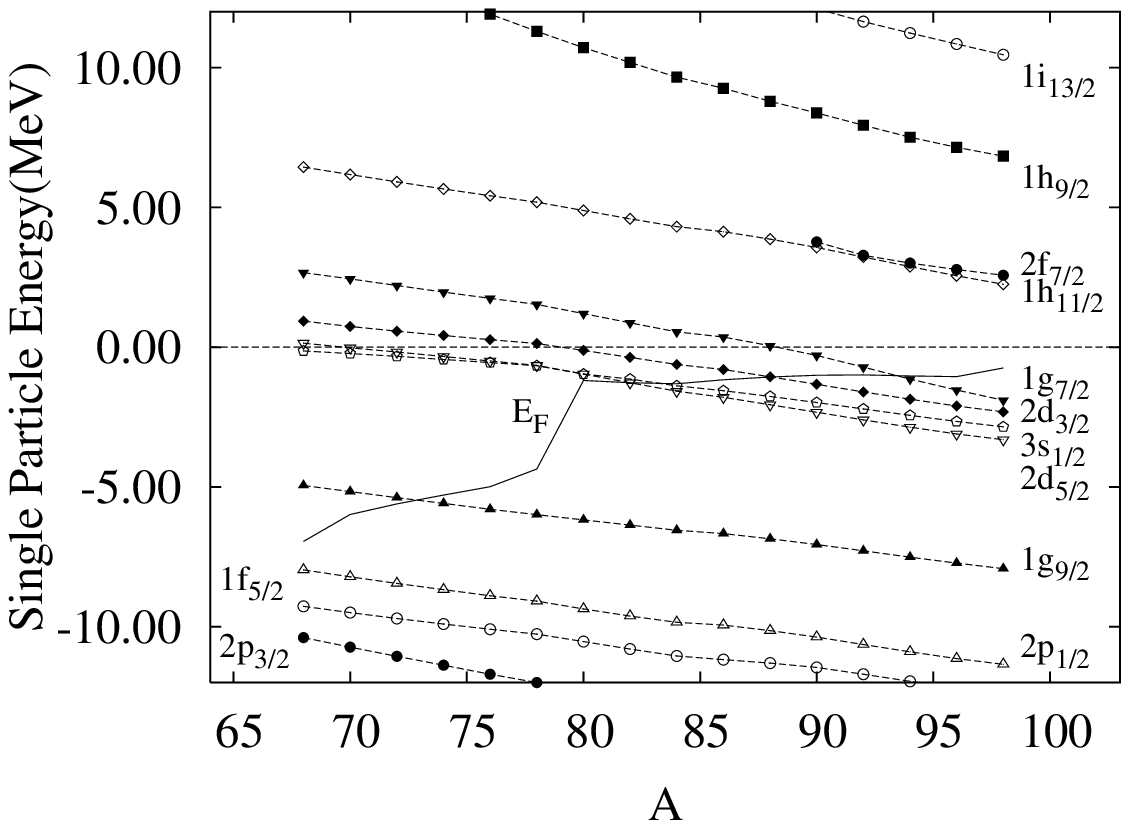}}
\caption{Single particle neutron energy levels in even
Ni isotopes. The Fermi level is shown by the continuous line.
\label{Nilev}}
\end{figure}

\end{document}